\documentclass[%
 reprint,
superscriptaddress,
twocolumn,
amsmath,amssymb,
prl,
floatfix,
]{revtex4-2}

\usepackage{verbatim}
\usepackage{comment}
\usepackage{graphicx}
\usepackage{dcolumn}
\usepackage{bm}
\usepackage{hyperref}

\usepackage{lineno} 

\usepackage{amssymb}

\usepackage[utf8]{inputenc}
\usepackage{ulem}
\usepackage{color,soul}
\usepackage{cancel}
\usepackage{siunitx}
\usepackage{xcolor}

\usepackage{xfrac}

\begin{document}

\newcommand{\aidan}[1]{{\color{red}#1}}

\newcommand{\nils}[1]{{\color{blue}#1}}

\newcommand{\ben}[1]{{\color{olive}#1}}


\title{Kinetics of electron–phonon scattering in silicon resolved by Rydberg transitions of donors}

\author{N. Dessmann}
\affiliation{Radboud University, Institute for Molecules and Materials, FELIX Laboratory, Nijmegen, The Netherlands}

\author{A. G. McConnell}
\affiliation{Paul Scherrer Institute,
Forschungsstrasse 111,
5232 Villigen,
Switzerland}
\affiliation{Laboratory for Solid State Physics, ETH Zurich, Zurich, CH-8093, Switzerland}
\affiliation{Quantum Center, ETH Zurich, Zurich, CH-8093, Switzerland}

\author{S. G. Pavlov}
\affiliation{Institute of Space Research, German Aerospace Center (DLR), Berlin, Germany}

\author{G. Matmon}
\affiliation{Paul Scherrer Institute,
Forschungsstrasse 111,
5232 Villigen,
Switzerland}

\author{G. Aeppli}
\affiliation{Paul Scherrer Institute,
Forschungsstrasse 111,
5232 Villigen,
Switzerland}
\affiliation{Laboratory for Solid State Physics, ETH Zurich, Zurich, CH-8093, Switzerland}
\affiliation{Quantum Center, ETH Zurich, Zurich, CH-8093, Switzerland}
\affiliation{Institut de Physique, EPF Lausanne, Lausanne, CH-1015, Switzerland}

\author{N. V. Abrosimov}
\affiliation{Leibniz-Institut für Kristallz{\"u}chtung (IKZ), Berlin, Germany}

\author{H. Paudyal}
\affiliation{Department of Physics and Astronomy, University of Iowa,
Iowa City, IA 52242
United States}

\author{M. E. Flatt\'e}
\affiliation{Department of Physics and Astronomy, University of Iowa,
Iowa City, IA 52242
United States}
\affiliation{Department of Applied Physics and Science Education, Eindhoven University of Technology, 5600 MB Eindhoven, The Netherlands}

\author{B. N. Murdin}
\email{b.murdin@surrey.ac.uk}

\affiliation{Advanced Technology Institute and Department of Physics, University of Surrey, Guildford, GU2 7XH, UK}

\date{\today}

\begin{abstract}

Rydberg states of atoms in vacuum are now well recognized as a resource for quantum technologies. Donors in semiconductors also display analogous states, which have been proposed for similar applications. While they benefit from permanent locations in their host crystals,  electron–lattice coupling leads to much shorter excited-state lifetimes than for neutral atoms in vacuum. Here we provide a quantitative description of donor–phonon kinetics, creating a basis for engineering donor systems in realistic material stacks for quantum devices.
Our theory incorporates both form factors for the Rydberg states, which given their large extents in real space provide strong selectivity in momentum space, and tabulated deformation potentials for all six phonon branches throughout the Brillouin zone. By confronting this framework with carefully controlled time-resolved free electron laser measurements, we show that the widely quoted position of the silicon conduction-band minimum, $k_0$, is inconsistent with observed donor relaxation rates and that quantitative agreement is obtained for a value further from the X-point than commonly assumed.
This stringent experiment–theory comparison establishes donor relaxation as a precision metrology for conduction band parameters and scattering processes in silicon, with consequences spanning from quantum devices to classical electronics.

\end{abstract}


\maketitle


Donor-bound electrons in silicon, long regarded as solid-state analogues of hydrogen atoms, provide a unique platform where atomic-scale quantum effects meet semiconductor technology \cite{murdin_nc_2013,Stemp2024,Edlbauer2025}. Their sharp level structure makes them exquisitely sensitive to lattice vibrations, turning them into precision probes of electron–phonon coupling. Understanding phonon-mediated relaxation in donors is not only central for quantum information schemes but also for benchmarking the microscopic scattering processes that underpin transport in classical silicon devices \cite{jacoboni1983,Li2021,Uchiyama2023}. Despite decades of effort using frequency-domain \cite{Jagannath1981a,Steger2009} and time-domain \cite{Vinh2008f,Lynch2010,Greenland2010,Huebers2013a,Litvinenko2014a,Chick2017,Dessmann2022} techniques, as well as nonlinear studies \cite{Pavlov2008}, a consistent picture of donor relaxation has remained elusive. Effective-mass and first-principles theories disagree on key rates \cite{Tsyplenkov2008a,Tsyplenkov2009,Zhukavin2020,Tyuterev2010}, and experiments themselves report lifetimes that differ by orders of magnitude. This persistent mismatch signals not a minor technicality but a fundamental gap in our understanding of how electrons and phonons interact in silicon at the  microscopic level.

\begin{figure*}
\includegraphics[width=0.95\linewidth]{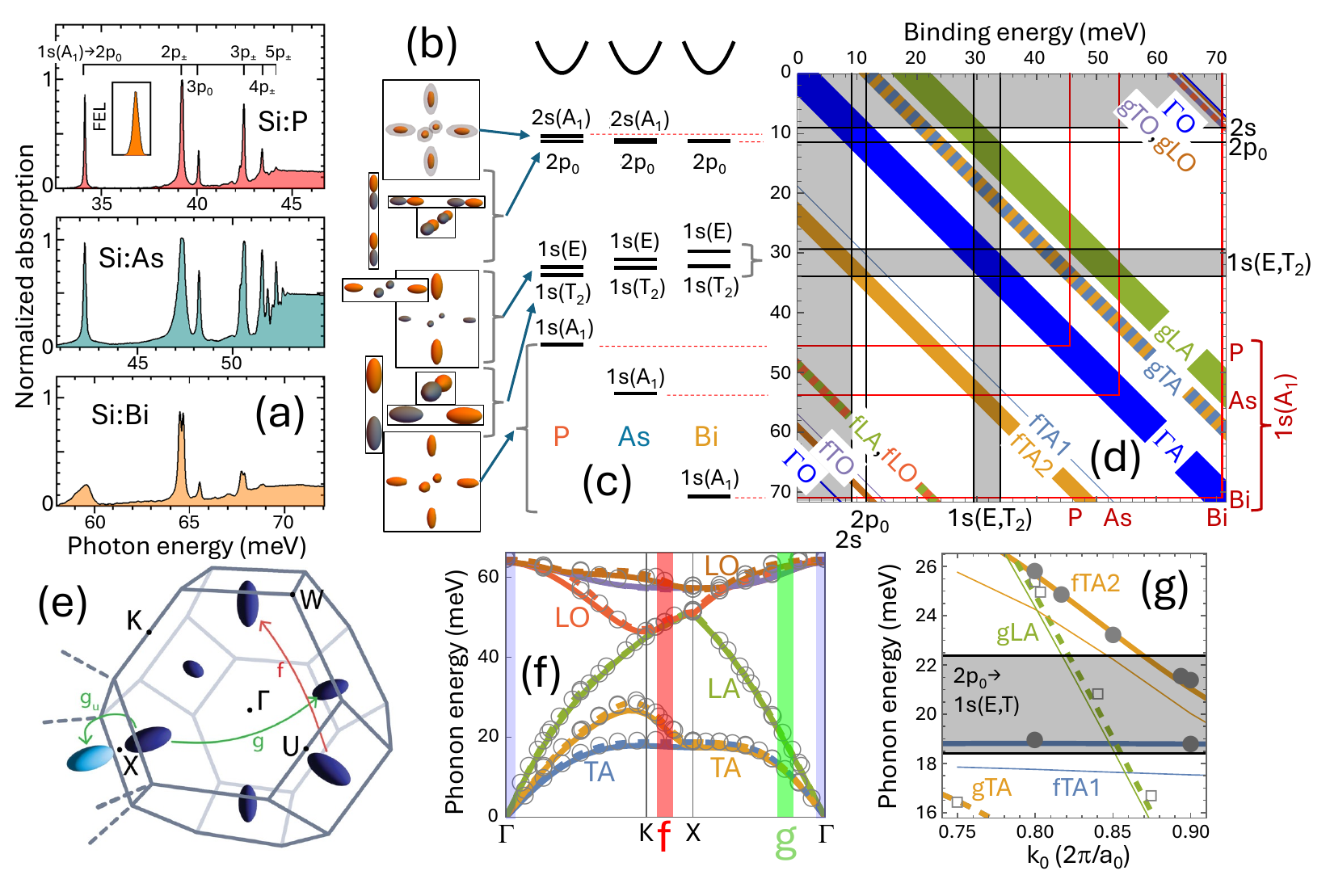}
\caption{
\textbf{g- and f-phonon relaxation in silicon donors.}
(a) Small-signal spectra of Si:P,As,Bi measured in transmission at $T=5$~K. Spectra are on the same energy scale but shifted, and all the transitions align.  A typical FELIX laser spectrum is shown as an inset: the pulse duration must be smaller than the relevant lifetime, which means the laser band width must be larger than the line width. 
(b) A selection of  relevant $k$-space wavefunctions  with sign indicated by the colour. 
(c) The species-dependent binding energies of the states in (b).  
(d)  Bound-state binding energy is indicated for excited states by black vertical and horizontal lines, and those for species specific ground states by red lines. Lines of constant transition energy are parallel to the diagonal, and those matching allowed g-phonon energies are in the upper triangle and f-phonons in the lower triangle. Colours correspond to panel (f). Intersection of vertical, horizontal and diagonal lines indicates a resonance. The widths of the diagonal phonon lines are proportional to the group velocity, which (along with the atomic form factor $\mathcal F(\Delta\mathbf{k})$) sets the resonance width. The gray bands indicate closely spaced states. Blue diagonals are the intravalley O and A phonons. The 2s energy indicated is the average of the A$_1$,E,T$_2$ states from effective mass theory.
(e) The first Brillouin Zone (BZ) with pockets centred on the c.b. minima  at  the $\mathbf{k}_0=k_0\langle100\rangle$ points. Allowed intravalley transitions involve g-type processes (green, between parallel valleys, $\Delta\mathbf{k}_\text{g}=2k_0\langle100\rangle$)  and f-type (red, perpendicular valleys, $\Delta\mathbf{k}_\text{f}=k_0\langle110\rangle$).
(f) The silicon phonon dispersion. Experimental neutron scattering data (\cite{Flensburg1999} and references therein) shown as symbols. 
The lines are from theory:  first principles (this work, solid lines), BvK model (dashed lines,   \cite{Flensburg1999}). The  g- and  f-phonons  are indicated for $k_0=0.81$. Increasing $k_0$ moves $\Delta\mathbf{k}_\text{g}$ towards $\Gamma$ and  $\Delta\mathbf{k}_\text{f}$ towards X. The transverse (T), longitudinal (L), acoustical (A) and optical (O) character of the branches is indicated. The colours of the lines are assigned in numerical order: the LO phonon is salmon at the f-point and brown at the g-point.  
(g) Phonon energies from (f) as a function of $k_0$, shown near the $2p_0 \!\to\! 1$s(E,T) transitions (grey band). Thick (thin) lines: BvK (first-principles) calculations; solid (dashed) lines: f- (g-) phonons. Filled (open) symbols show experimental f- (g-) data from panel (f).
}
\label{fig:ftir_ladder}
\end{figure*}

A central parameter in calculations of electron-phonon scattering in silicon is the reciprocal-space location of the conduction-band minimum, $k_0$ \cite{jacoboni1983,Li2021,Uchiyama2023}. This wavevector selects the phonon momenta, and hence the phonon energies \cite{Flensburg1999}, available for intervalley scattering contributions, termed g- and f-phonons depending on whether the valleys involved have parallel or perpendicular axes, respectively.
Because the most efficient relaxation channel for silicon donors is the emission of single g/f-phonons \cite{Tsyplenkov2008a,Tsyplenkov2009}, and the sharp energy spectrum of donor Rydberg states strongly restricts the available decay pathways through energy conservation Fig \ref{fig:ftir_ladder}, $k_0$ directly controls donor relaxation dynamics.
At the same time, $k_0$ renders electron wavefunctions oscillatory producing interference that sets the strength of qubit-qubit interactions in proposed readout and gating schemes \cite{gamble2015,Stemp2024,Steinacker2025,Edlbauer2025}. Yet despite its importance, the accepted value of $k_0$ has remained  uncertain.

The widely quoted value $k_0 = 0.85$ (in units of $2\pi/a_0$, where $a_0$ is the lattice constant) originates from early tight-binding fits to band-edge energies and effective masses, reinforced by comparisons with the oscillatory super-hyperfine patterns of Si:P \cite{Feher1959,Ivey1975a}. Later tight-binding models, however, favoured smaller values closer to 0.75 \cite{Klimeck2000}, while optical probes of intervalley phonons suggested intermediate values around 0.79-0.82 \cite{Macfarlane1958,Dumke1960,Onton1969,Folland1970}. A magneto-stress-transport measurement \cite{Eaves1974} yielded a higher value, but required unexplained breaking of selection rules in its interpretation. The conflicting evidence has left the field relying on a number that is often repeated, but rarely scrutinized. By confronting donor relaxation lifetimes with established phonon dispersions \cite{Kulda1994,Flensburg1999} and new first-principles deformation potentials, we provide a decisive test of which values of $k_0$ are consistent with experiment.

Although more direct experimental probes of the conduction-band valley position in silicon have recently become available, these measurements have primarily focused on heavily doped, near-surface systems \cite{Holt2020, Voisin2020, Constantinou2023}. Scanning probe studies of individual donors reveal real-space wavefunctions whose Fourier spectra peak at $k_0 \approx 0.81$ \cite{Voisin2020}, while more bulk sensitive soft X-ray angle-resolved photoemission spectroscopy (SX-ARPES) measurements of $\delta$-doped silicon samples yield values in the range $k_0 \approx 0.81$ \cite{Constantinou2023}. The mutual consistency of surface- and delta layer-sensitive experiments supports the view that the commonly quoted bulk value $k_0 \approx 0.85$ may be an overestimate. However, because the measurements are performed in strongly perturbed electronic environments (including heavy doping, quantum confinement, and proximity to surfaces or interfaces) a direct, bulk-sensitive determination of $k_0$ in lightly doped silicon has remained unavailable.

Here we use relaxation between excited Rydberg states in donors as a bulk probe of electron–phonon interactions. Crucially, we combine several complementary time-resolved THz techniques and carefully control both temperature and excitation density. Alongside an improved theoretical framework incorporating full-branch deformation potentials, this allows quantitative comparison across multiple donor species. In this way, we show that a long-standing factor-of-four discrepancy between experiment and theory for the rates disappears when $k_0$ is shifted very modestly from 0.85 to 0.81, establishing donor dynamics as a sensitive benchmark for this key band-structure parameter.

\begin{figure}
\includegraphics[width=0.95\linewidth]{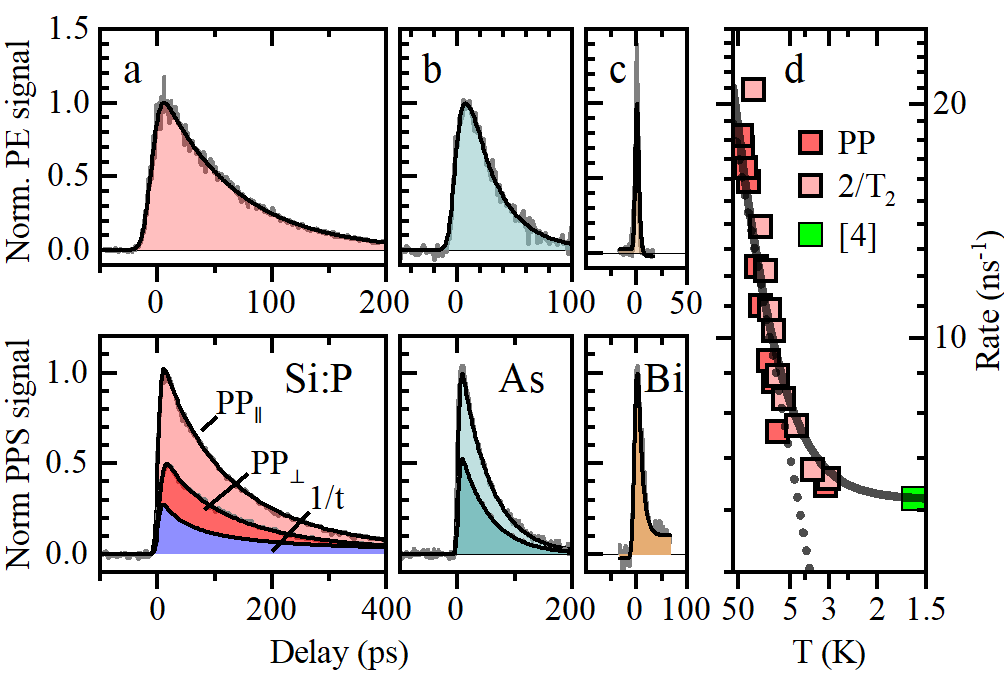}
\caption{
\textbf{Experimental dynamics.}
(a) Time-resolved  spectroscopy at base temperature ($T=3$~K) and low intensity for  Si:P.  Note that the echo decay $T_E=T_2/4$ \cite{Greenland2010} and ideally $T_1=\frac12T_2=2T_E$. The normalized PE (top row) and PP$_{\parallel,\perp}$ (bottom row) are shown with ordinate axes scaled by a factor of 2. Shown superimposed on each PE is a single exponential (with an error-function rise at $t=0$). Shown on the PP (bottom row) is the sum of a single exponential and a reciprocal decay (blue) such that the reciprocal is common to both PP$_{\parallel,\perp}$. (b,c) Same as (a) for Si:As and Si:Bi, with the same abscissa scale. No reciprocal component was apparent in these cases, and for Si:Bi only PP$_{\perp}$ was obtained. (d) Temperature dependence  for Si:P (on log-reciprocal axes). Also shown is $2/T_2$ from frequency domain \cite{Steger2009} (green). The dashed line shows an activated process with $\Delta E/k_B=10$K. 
} 
\label{fig:PPES_PASBi}
\end{figure}

Fig \ref{fig:ftir_ladder}, shows the six-fold degenerate hydrogenic orbitals of interest, whose Fourier components are concentrated at the conduction band minima.
The central cell correction (a.k.a. the quantum defect or chemical shift) of the donor ion lifts the degeneracy and mixes the valley contributions to produce states of character A$_1$ (a singlet), E (a doublet) and T$_2$ (a triplet).  
At low temperature, the lowest energy optically allowed transition is 1s(A$_1)\to2$p$_0($T$_2)$. 
Population relaxation from the excited state proceeds by phonon emission. To contribute, the phonon must simultaneously satisfy energy and momentum conservation, shown when vertical, horizontal and diagonal lines all intersect on Fig \ref{fig:ftir_ladder}d. The phonon must, of course, also produce a large perturbation energy known as the deformation potential.
In the case of Si:P the direct relaxation 2p$_0($T$_2)\to1$s(A$_1)$ is far from any resonance, but 2p$_0($T$_2)\to1$s(E,T$_2)$ is close to several, depending on $k_0$, Fig \ref{fig:ftir_ladder}g. 

For a transition from $\lvert a \rangle \to \lvert b \rangle$, the Golden Rule for the rate is \cite{Tsyplenkov2008a}:
$$W_{ab}= \frac{2\pi}{\hbar} \sum_{\mu} \int d^3q ~ |\mathcal{M}_{ab\mu}  (\mathbf{q})|^2 ~ \delta(E_{ab}-\hbar\omega_{\mu}(\mathbf{q})),$$
where  $E_{ab}$ is the transition energy and $\hbar\omega_{\mu}(\mathbf{q})$ is the phonon energy for branch $\mu$  at wavevector $\mathbf{q}$. 
The wavefunctions are products of a slowly varying hydrogenic envelope function and a lattice-periodic Bloch part, and the matrix element may be correspondingly separated:
$ \mathcal{M}_{ab\mu} (\mathbf{q})  = (16\pi^3\rho E_{ab})^{-1/2} \, \mathcal D_{\mu }(\mathbf{q})\mathcal{F}_{ab}(\mathbf{q})  .$
Here the $D_{\mu}(\mathbf{q})$ is the deformation potential arising from the quickly oscillating part and the ion displacement within each cell, 
and $\rho$ is the crystal mass density.  $\mathcal{F}_{ab}(\Delta \mathbf{k})=\langle{a}|e^{i\Delta \mathbf{k}\cdot\mathbf{r}}|b\rangle$ is known as the form factor. 
It is the Fourier transform of the product of the two envelope functions, or equivalently the convolution of the reciprocal-space wavefunctions from Fig \ref{fig:ftir_ladder}b. 
The $k$-space convolution is only non-zero for momentum transfer $\Delta \mathbf{k}$ near $\Gamma$, f or g (Fig\ref{fig:ftir_ladder}e).
In the approximation that the donor wavefunction envelopes are anisotropic hydrogen-like functions and for the case of transitions involving  1s via g- or $\Gamma-$phonons (so that valleys are parallel), $\mathcal{F}_{ab}(\Delta\mathbf{k})$ may be written in terms of hypergeometric functions (the Bethe-Gordon formula) \cite{Meremianin2006}. For f-phonons or higher order transitions simple scaling rules are not  available.

Energy conservation is strictly enforced by the Dirac $\delta$-function, while momentum conservation is produced more softly by $\mathcal{F}$. The finite width of its features at $\Gamma$,f,g embodies the Heisenberg uncertainty arising from the finite spatial extent of the states.
Thus, if an electronic transition and a  phonon  have an energy detuning of $\delta E\gtrsim \hbar v_g \delta k$ where $v_g$ is the phonon group velocity and  $\delta k$ is the width of $\mathcal F$ then the process is forbidden, as shown on Fig \ref{fig:ftir_ladder}d. 
Indeed the simultaneous momentum-space restriction  and energy conservation is what makes donor relaxation such an exquisite probe of $\mathcal D$: it is selective enough to isolate specific processes, yet permissive enough that relaxation proceeds at measurable rates.

To extract population relaxation times from frequency-domain spectroscopy, extremely clean samples are required and the experimental conditions must be carefully controlled to eliminate inhomogeneous broadening. The sharpest donor lines in silicon have been obtained using residual impurities in undoped, isotopically enriched \textsuperscript{28}Si, with the sample immersed in superfluid He in a strain-free mount \cite{Steger2009}. In that study, the 1s(A$_1)\to2$p$_0$ resonance of Si:P was Lorentzian, consistent with homogeneous broadening, giving a dephasing time $T_2 = 1/\pi\Delta f = 320$~ps where $\Delta f$ is the full width at half maximum, and hence a presumed population decay time of $T_1 = T_2/2 = 160$~ps.

In the time domain, the most common technique for measuring $T_1$ is single-color pump–probe (PP) spectroscopy, which has reported values ranging from 140–235~ps for 1s(A$_1)\to2$p$_0$ in Si:P \cite{Vinh2008f,Lynch2010,Huebers2013a,Litvinenko2014a}. These are close to the frequency-domain estimate but show considerable variation. Photon echo (PE) spectroscopy, which measures $T_2$ directly, has been applied to Si:P and reported $T_2 \approx 160$~ps \cite{Greenland2010}, far below the small-signal value. It was suggested \cite{Vinh2008f,Greenland2010} that high excitation intensity causes parasitic two-photon excitation into the conduction band, leading to disagreement. At sufficiently low temperature and intensity, however, the conduction band does not play a role and the dynamics reduce to those of the six 1s and six 2p$_0$ states (Fig.~\ref{fig:ftir_ladder}).

We investigated Si  doped with P, As, or Bi, because of the wide variation in level structure, Fig \ref{fig:ftir_ladder}. Float-zone-grown  [100] crystals were used, with concentrations 
$1.28, 3.6, 2.7\times10^{15}$cm$^{-3}$ respectively.
Isotopically enriched $^{28}$Si samples were used where available (P and Bi), but it is not clear that this is necessary.
From the small-signal spectrum, Fig \ref{fig:ftir_ladder}, none of the samples  exhibits a Lorentzian  1s(A$_1)\to2$p$_0$ line:  in the P and As samples concentration broadening dominates while in Si:Bi both lifetime and inhomogeneous broadening contribute.

We performed  time-resolved experiments with light from the free electron laser FELIX (Fig \ref{fig:PPES_PASBi}). The experiment combines PP and  PE spectroscopy in a single setup with independent polarization control of the different beams.  The details of the set-up are described elsewhere \cite{Dessmann2022}.  
Far-infrared beams have a large waist and divergence, and the use of multiple beams requires a large numerical aperture. This precludes the immersion cryostat geometry employed for small-signal transmission spectroscopy \cite{Steger2009}. Instead, samples were mounted on a cold finger in vacuum, with a silver-paste ring ensuring good thermal contact while minimizing strain.

Particular care was taken to reduce systematic errors. Beam profiles, overlap, and focusing were optimized using automated IR imaging with a pyroelectric camera, which enabled high signal-to-noise and low scattered-light background. This allowed us to operate at much lower intensity than in previous time-resolved work \cite{Vinh2008f,Lynch2010,Huebers2013a,Litvinenko2014a,Greenland2010}. Only scans that were stable and repeatable without drift were retained, and all data presented here are unprocessed apart from normalization.

Because of the electric dipole selection rules, only one of the 2p$_0$(T$_2$) states is pumped. The PE signal therefore decays at a rate equal to the sum of all relaxation channels out of this single state.
In a PP experiment the transient absorption is proportional to the population difference between the ground and the probed excited state. PP$_\parallel$ (parallel polarized beams) directly measures the emptying of the pumped state. It is also sensitive to any back-filling from other 2p$_0$ components (which the PE does not detect) and to refilling of the ground state.
PP$_\perp$ instead probes a different 2p$_0$(T$_2$) state that is initially unoccupied. Its population difference starts at half the value of PP$_\parallel$, and it reports the scattering among the 2p$_0$ states as well as refilling of the ground state.
Figure~\ref{fig:PPES_PASBi}a shows the case of Si:P. Two clear features emerge: (i) the difference between PP$_\parallel$ and PP$_\perp$ decays with half the coherence time $T_2$ obtained from the photon echo, and (ii) the two PP signals do not converge until the decay is complete.
This behavior has a straightforward interpretation. First, donors must spend essentially no time in the intermediate 1s(E,T$_2$) states - either because they are bypassed by direct relaxation to the 1s(A$_1$) ground state, or because 1s(E,T$_2$)$\to$1s(A$_1$) relaxation is extremely fast. In both cases, the  time scales for emptying the excited state and refilling the ground state are identical. Second, scattering among the excited-state components must be very slow, so there is no rapid equalization of populations.
A  rate-equation model confirms that this is the only consistent scenario. We can therefore identify $T_1$ directly from the PP$_\parallel$–PP$_\perp$ difference (not shown), with $T_2 = 2T_1$, exactly as for an ideal two-level atom.

There is also a reciprocal component in the PP signals,  even at the lowest intensity used, indicative of two-photon ionization and recombination \cite{Vinh2008f}. This affects both PP$_{\parallel}$ \& PP$_{\perp}$ equally, and does not influence the difference. We do not consider it further in this work. 

The Si:P data of Fig \ref{fig:PPES_PASBi}a were taken with the lowest possible temperature and intensity. Substantially lower intensity was possible for the  PE compared with the PP (a factor of 10), because the former is background-free. At the lowest  temperature  $2/T_2$ and the PP$_\perp$ rate both saturate, as shown  in Fig \ref{fig:PPES_PASBi}d. 
The temperature dependence above saturation is consistent with an activated process of energy $\Delta E/k_B = 10$ K, which corresponds closely to the 2p$_0 \to 2$s(A$_1$) transition energy of 0.87~meV (Fig.~\ref{fig:ftir_ladder}).

For Si:As, we reduced the intensity sufficiently to observe $T_2 = 2T_1$ as shown in Fig \ref{fig:PPES_PASBi}b, meaning that the system is effectively a two-level system, the decay time did not saturate around base temperature, i.e. it is significantly more temperature sensitive than Si:P. We note that the  2p$_0\to2$s(A$_1)$ energy is much smaller, 0.22~meV (Fig \ref{fig:ftir_ladder}), which corresponds to an activation temperature of only 2.5~K -- below our experimental base temperature.

For Si:Bi (Fig.~\ref{fig:PPES_PASBi}c) we also reached sufficiently low temperature and intensity to saturate the lifetime. In this case the relaxation is much faster than for either P or As.

\begin{figure}[b]
\includegraphics[width=\linewidth]{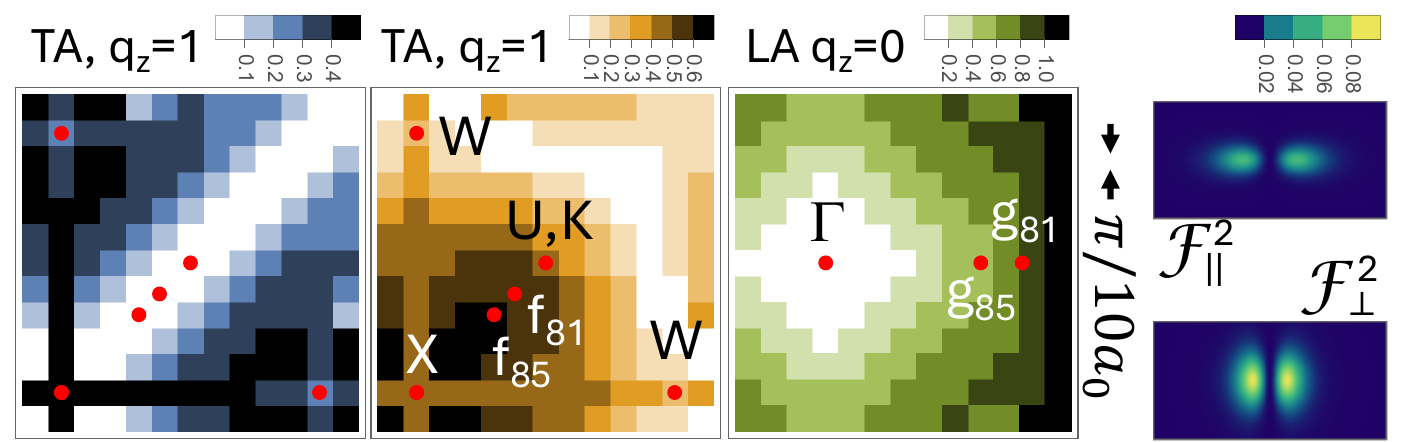}
\caption{
\textbf{The matrix element $\mathcal M\propto \mathcal D \mathcal F$.} 
The deformation potential, $\mathcal D_\mu$, in eV/{\AA}, for the two f-TA branches and the g-LA branch. Colours correspond to Fig \ref{fig:ftir_ladder}f.  
High symmetry points are indicated, along with $\Delta\mathbf{k}_\text{g,f}$ for $k_0=0.81$ and 0.85.
The (dimensionless) contributions to the form factor $|\mathcal{F}|^2$ near $\Delta\mathbf{k}_\text{g,f}$ are shown for 2p$_0\to1$s(E,T) are also shown. The scale bar indicating $\frac1{20}{2\pi}/{a_0}$ is the same for all panels.}
\label{fig:defpot}
\end{figure}

To compare these experimental lifetimes  with the Golden Rule,  we carried out first-principles calculations for  $\hbar\omega_\mu(\mathbf{q})$ and $\mathcal D_\mu(\mathbf{q})$ using the Quantum ESPRESSO package \cite{Giannozzi2017}. Relativistic norm-conserving pseudopotentials \cite{vanSetten2018} were employed in conjunction with the Perdew–Burke–Ernzerhof exchange-correlation functional within the generalized gradient approximation \cite{Perdew1996}. A $\Gamma$-centered $12 \times 12 \times 12$ Monkhorst–Pack $k$-point mesh, a plane-wave energy cutoff of 80 Ry (with a charge density cutoff of 320 Ry), and a smearing parameter of 0.01 Ry were used for BZ sampling. 
$\hbar\omega_\mu(\mathbf{q})$ and $\mathcal D_\mu(\mathbf{q})$ were extracted from the electron–phonon interaction matrix elements computed using density functional perturbation theory, which evaluates the 1$^\text{st}$-order variation of the self-consistent potential with respect to atomic displacements along phonon eigenmodes.

The phonon dispersion is shown in Fig.~\ref{fig:ftir_ladder}f. 
Despite having no adjustable parameters, the first-principles calculation reproduces experimental neutron-scattering data very well, and is almost as accurate a theoretical description of this as a Born–von Karman (BvK) model with  21 fitting parameters \cite{Flensburg1999}. 
For the rate calculations  we used the BvK model, since it offers slightly better accuracy for f-type phonons, and resonant BvK values are shown vs $k_0$ in Table \ref{tab:phononfreqs}. 
We note, however, that improved agreement of first-principles dispersion with neutron data has been reported using a different functional \cite{Hummer2009}.  
We further note that the energy surfaces defined by the $\delta$-function appearing in $W_{ab}$ are planar for g-phonons to a good approximation, in agreement with earlier assumptions \cite{Tsyplenkov2008a}, but very strongly curved for f-phonons.

The 2p$_0 \to 1\text{s(E,T}_2)$ transitions (Fig.~\ref{fig:ftir_ladder}g) all lie close to 20~meV. 
From the phonon dispersion (Fig.~\ref{fig:ftir_ladder}f,g) we see that at this energy only a few phonon relaxation pathways are relevant: f-TA and g-LA. 
Slices of the appropriate deformation potentials $\mathcal D_\mu(\mathbf{q})$ for these cases are shown in Fig.~\ref{fig:defpot}. Resonant values are in Table \ref{tab:phonon-coupling-transposed-set2} and a full tablulation across the BZ is in Supplementary Data. 
These results agree well with earlier partial datasets, some of which averaged over multiple branches \cite{jacoboni1983,Tyuterev2010,Tang2012,Li2021}. 
Figure~\ref{fig:defpot} also illustrates that $\mathcal D= 0$ along [110] for the lower f-TA branch, as required by symmetry \cite{Tyuterev2010}, while $\mathcal D_{\mathrm{TA2}}(\Delta\mathbf{k}_\text{f})\approx 0.6$~eV/\AA{} for the upper branch. 
Because scattering rates scale as $D^2$, and the two branches differ strongly in frequency, averaging across them - as was done in some earlier donor relaxation \cite{Tsyplenkov2008a}, spin lifetime \cite{Tang2012} and mobility studies \cite{jacoboni1983,Li2021} - can obscure the dominant contribution and risks  qualitatively incorrect conclusions. 

To complete the matrix element $\mathcal M$ we calculated the  form factor $\mathcal F$ for 2p$_0\to1$s near $\Delta\mathbf{k}_\text{g,f}$, Fig.~\ref{fig:defpot}.  
We used anisotropic hydrogenic envelopes \cite{Tsyplenkov2008a} with updated silicon band-edge effective masses \cite{Li2018}. 
It can be seen that $\mathcal F(\Delta\mathbf{k})$ near $\Delta\mathbf{k}_\text{f}$ is somewhat more compact and greater in height than $\Delta\mathbf{k}_\text{g}$, meaning that f-phonon resonances are stronger and sharper for a given $\mathcal D$ and $v_g$.

\begin{figure}[b]
\includegraphics[width=0.95\linewidth]{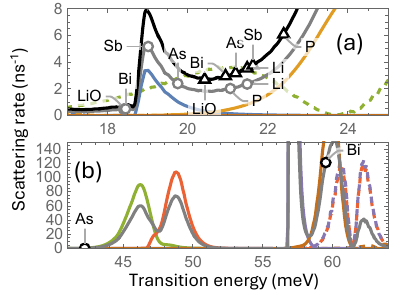}
\caption{
\textbf{Theoretical relaxation rates.}
(a) Calculated  intervalley relaxation  ${W}_{ab}$ for the case that $a\to b$ is 2p$_0$(T$_2)\to1$s(E) (grey line) and 1s(T$_2$) (black line), assuming $k_0=0.81$.  These multi-valley rates are found from a weighted sum of single-valley rates for different phonon branches, shown as coloured lines.  Colours indicate the branch, corresponding to Figs \ref{fig:ftir_ladder}f: dashed lines: g-phonons, solid lines: f-phonons. Predicted rates for specific species are shown with symbols. (b) Same as (a) for 2p$_0$(T$_2)\to1$s(A$_1$) (grey line).}
\label{fig:rate}
\end{figure}

Figure~\ref{fig:rate}a shows the rates for 2p$_0$(T$_2)\to1$s(E) and 1s(T$_2$).  
The  lower f-TA branch has very low group velocity and a very small deformation potential, giving a sharp, weak contribution peaked  at 18.8~meV (blue). The low $v_g$ also makes the resonance almost independent of $k_0$.  
The upper f-TA branch, with both large $\mathcal D$ and higher velocity gives a strong, broad feature (orange) peaked  at 25.21~meV for $k_0=0.81$.  
Earlier first-principles work predicted much longer lifetimes, possibly because only the lower f-TA branch was included \cite{Tyuterev2010}, while earlier effective mass theory predicted much higher rates, primarily because of an assumed f-TA phonon resonance associated with an inappropriate $k_0$.

Figure~\ref{fig:rate}b shows the relaxation rates to the 1s(A$_1$) state, which is only relevant for Bi and As. 
For Bi, the transition lies in the middle of the TO resonances. 
The total predicted rates for 2p$_0$(T$_2)\to1$s(A$_1$,E,T$_2$) are given in Table \ref{tab:dopant-vs-k0}. 
For P and Bi the agreement with experiment (also Table \ref{tab:dopant-vs-k0}, taken from Fig \ref{fig:PPES_PASBi}) is good, while for As the lack of saturation at low temperature limits the comparison.  

For the 2s(A$_1)\to1$s(A$_1$,E,T$_2$) transitions we obtain 29 and 24~ns$^{-1}$ for P and As, much faster than for relaxation out of 2p$_0$. 
This enables  the observed thermal activation of a rapid channel. 
The higher rate reflects the stronger overlap between 1s and 2s wavefunctions, and hence a larger form factor. 
A large $\mathcal{F}$ is also responsible for very fast  relaxation from 1s(E,T$_2)\to$1s(A$_1$) in P and As (40 and 100~ns$^{-1}$, respectively), via f-TA and g-TA phonons. 

Transitions involving phonons with energy less than about 10~meV can only be produced by intravalley $\Gamma$ phonons (Fig.~\ref{fig:ftir_ladder}d), where  $\mathcal D$ necessarily vanishes, since a long-wavelength phonon corresponds to a rigid translation of the lattice and produces no local deformation (as shown in Fig \ref{fig:defpot} for the case of LA near $q=0$). 
As a result, transitions within the excited-state manifold, or among the 1s(E,T$_2$) states, are intrinsically weak, contrary to earlier claims \cite{Tsyplenkov2008a}. 
The observed thermal activation from 2p$_0\to2$s(A$_1$) must therefore involve a different mechanism, most likely inelastic scattering by free electrons, or possibly two-phonon processes, which we do not analyze further here.

The P data in Fig.~\ref{fig:rate} fall near  the steep rise associated with the f-TA resonance. Increasing $k_0$ brings this resonance down in energy rapidly due to the large group velocity (Fig.~\ref{fig:ftir_ladder}g). 
A change of only 5\% in $k_0$, from 0.81 to 0.85, shifts the f-TA resonance almost directly onto the Si:P 2p$_0\to1$s(T$_2$) transition, increasing the predicted rate by a factor of four (Table \ref{tab:dopant-vs-k0}) -- well above both our experimental result and the frequency-domain value. 
Our results therefore strongly disfavour $k_0 \approx 0.85$.

In conclusion, donor relaxation provides a uniquely stringent test of electron–phonon interaction theory in bulk silicon.
By combining carefully controlled time-resolved spectroscopy with established phonon dispersions and enhanced first-principles calculations, we resolve long-standing discrepancies in donor relaxation between experiment and theory, finding quantitative agreement when the conduction-band minimum is set at $k_0 \approx0.81$, in agreement with recent direct measurements at surfaces \cite{Voisin2020} and for delta layers \cite{Constantinou2023}.

The implications extend beyond donor relaxation. The position of the conduction-band minimum governs the valley interference encoded in  electron wavefunctions, which in turn controls intervalley scattering rates relevant to carrier mobilities \cite{jacoboni1983,Li2021} and spin and valley relaxation \cite{Tang2012}, as well as oscillatory exchange and tunnel couplings between localized electrons in silicon quantum devices based on donors  \cite{gamble2015,Voisin2020,Stemp2024,Edlbauer2025} and dots \cite{Steinacker2025}. The same valley interference underlies the spatially oscillatory hyperfine coupling between a donor electron and neighbouring nuclear spins that historically informed early estimates of $k_0$  \cite{Feher1959,Ivey1975a}, and that is now exploited in multi-nuclear-spin quantum registers in silicon, including recent demonstrations of electron-mediated coupling between large nuclear-spin registers \cite{Edlbauer2025}. Our results show that even modest shifts in $k_0$ can produce order-unity changes in selected scattering channels, providing a microscopic explanation for the sensitivity observed here in Rydberg-state decay kinetics and establishing donor relaxation as a precision benchmark for electron–phonon processes in silicon.

\begin{table}[ht]
\centering
\begin{tabular}{l|cccccc}
$k_0$     & 0.75 & 0.77 & 0.79 & 0.81 & 0.83 & 0.85 \\
\hline
g-TA     & 16.47 & 15.83 & 15.06 & 14.16 & 13.12 & 11.94 \\
g-LA     & 30.26 & 28.14 & 25.96 & 23.71 & 21.40 & 19.03 \\
g-TO     & 60.01 & 60.48 & 60.96 & 61.45 & 61.94 & 62.41 \\
g-LO     & 62.27 & 62.69 & 63.05 & 63.36 & 63.62 & 63.83 \\
f-TA (1) & 18.81 & 18.81 & 18.82 & 18.82 & 18.81 & 18.81 \\
f-TA (2) & 27.60 & 26.91 & 26.10 & 25.21 & 24.26 & 23.30 \\
f-LA     & 44.91 & 45.58 & 46.25 & 46.90 & 47.54 & 48.17 \\
f-LO & 47.11 & 47.39 & 47.77 & 48.21 & 48.68 & 49.17 \\
f-TO (1) & 57.21 & 57.15 & 57.08 & 57.03 & 56.98 & 56.94 \\
f-TO (2) & 60.67 & 60.40 & 60.10 & 59.78 & 59.43 & 59.05 \\
\end{tabular}
\caption{Phonon energies, $\hbar\omega_\mu(\mathbf{q})$ (meV) for g- and f- scattering for all possible  phonon branches as a function of $k_0$. Results from the BvK model of Fig \ref{fig:ftir_ladder}f.}
\label{tab:phononfreqs}
\end{table}

\begin{table}[ht]
\centering
\begin{tabular}{l|cccccc}
$k_0$     & 0.75 & 0.77 & 0.79 & 0.81 & 0.83 & 0.85 \\
\hline
g-TA     & 0.24 & 0.27 & 0.30 & 0.31 & 0.31 & 0.30 \\
g-LA     & 1.32 & 1.15 & 0.99 & 0.83 & 0.69 & 0.55 \\
g-TO     & 3.48 & 3.49 & 3.50 & 3.52 & 3.53 & 3.55 \\
g-LO     & 0.00 & 0.00 & 0.00 & 0.00 & 0.00 & 0.00 \\
f-TA (1) & 0.00 & 0.02 & 0.02 & 0.01 & 0.02 & 0.00 \\
f-TA (2) & 0.46 & 0.51 & 0.55 & 0.58 & 0.60 & 0.61 \\
f-LA     & 1.95 & 2.07 & 2.20 & 2.02 & 1.10 & 0.49 \\
f-LO     & 0.86 & 0.79 & 0.72 & 1.33 & 2.28 & 2.60 \\
f-TO (1) & 4.58 & 4.58 & 4.58 & 4.59 & 4.59 & 4.59 \\
f-TO (2) & 2.83 & 2.74 & 2.63 & 2.50 & 2.36 & 2.19 \\
\end{tabular}
\caption{Phonon deformation potentials  $\mathcal D_\mu(\mathbf{q})$ (in eV/{\AA}) for all possible  phonon resonances as a function of $k_0$. Results are interpolations of the dataset in Fig \ref{fig:defpot} and Supplementary Data. Both g-LO and f-TA1 should have $\mathcal D$ equal zero, and discrepancies  are an indication of the possible error introduced by interpolation. }
\label{tab:phonon-coupling-transposed-set2}
\end{table}

\begin{table}[ht]
\centering
\begin{tabular}{l|rrrrrr|c}
       $k_0$ & 0.75 & 0.77 & 0.79 & 0.81 & 0.83 & 0.85 & Exp. \\
\hline
Li      & 1.68  & 2.58  & 4.99  & 7.59  & 10.67 & 31.93 &      \\
LiO     & 5.91  & 4.05  & 3.97  & 4.72  & 4.15  & 6.30  &      \\
P       & 2.56  & 3.02  & 5.17  & 8.04  & 18.28 & 33.99 &   7.7   \\
As      & 6.34  & 3.91  & 4.31  & 5.84  & 6.70  & 16.33 &   19   \\
Sb      & 5.19  & 6.93  & 8.21  & 9.57  & 12.11 & 24.38 &      \\
Bi      & 53.82 & 88.41 & 115.06 & 122.16 & 112.27 & 105.58 &  125    \\
\end{tabular}
\caption{Transition rates for 2p$_0$(T$_2)\to1$s(A$_1$,E,T$_2$) (ns$^{-1}$) for different dopants as a function of $k_0$. Experimental values shown are  $1/T_1$ values from fitting exponential decays on Fig \ref{fig:PPES_PASBi}. The uncertainty in the fits is smaller than the precision quoted, and the dominant error is likely to be the systematic uncertainty due to incomplete saturation in the limit of low temperature and intensity - this is small in the cases of P and Bi but very large for As. The limiting value for Si:P at low temperature is 6.2ns$^{-1}$ \cite{Steger2009}.}
\label{tab:dopant-vs-k0}
\end{table}

\begin{acknowledgments}
We gratefully acknowledge support from EPSRC-UK Grant number EP/M009564/1. H. P. and M. E. F. acknowledge support from AFOSR Award No. FA9550-24-1-0355 for {\it ab initio} calculations of electron-phonon coupling and phonon dispersion relations, while the work of A.G.M., G.M. and G.A. was supported by funding from the European Research Council under the European Union’s Horizon 2020 research and innovation program, within the Hidden, Entangled and Resonating Order (HERO) project with Grant Agreement 810451.
\end{acknowledgments}

\clearpage

\bibliography{clean_refs.bib}

\clearpage

\typeout{}

\end{document}